\documentclass[prl,twocolumn,superscriptaddress,showpacs,amsmath,amssymb,floatfix]{revtex4}
\usepackage{array}
\usepackage{mathdots}
\usepackage{graphicx,epsfig,bm,float,subfigure}
\usepackage{tabularx}
\usepackage{longtable,color,epsfig}

\DeclareRobustCommand{\orderof}{\ensuremath{\mathcal{O}}}

\begin{document}

\title{Mode folding in systems with local interaction: unitary and non-unitary transformations using
tensor states}

\author{Jose Reslen}

\affiliation{Coordinaci\'on de F\'{\i}sica, Universidad del Atl\'antico, Kil\'ometro 7 Antigua v\'{\i}a a Puerto Colombia, A.A. 1890, Barranquilla, Colombia.}


\begin{abstract}
An approach to the simulation of locally interacting systems  
is demonstrated and assayed. The proposal is built upon the
concept of folding of bosonic modes previously introduced 
in the context of linear dynamics and can be seen as an
alternative to Trotter-Susuki expansion in studies of 
quantum propagation based on tensor states. It is shown that 
evolution as well as ground state computations can be
implemented and that test simulations deliver comparatively 
accurate results. The whole analysis provides
insight into the way well-known quantum precursors affect
mean values and fluctuations in realistic setups.
\end{abstract}

\maketitle

\section{Introduction}
The study of numerical methods of quantum mechanics has
become an exciting and increasingly active field of research,
not only because numerical methods are extremely important
as simulation instruments, but also because they help
understand the underlying functioning of the quantum theory.
Such is the case of Density Matrix Renormalization Group 
(DMRG), which has played an important role in this sense 
since its inception. Time Evolving Block Decimation 
(TEBD) has appeared more recently and integrates elements
of DMRG with concepts of quantum computation. Both DMRG
and TEBD have inspired a number of variants such as tDMRG, iTEBD,
MPS, MERA and PEPS, to mention just a few \cite{Ulrich,ROrus}. However, 
there exist instances where the lack of efficient simulation 
protocols is yet an issue. Here, a proposal in this direction 
is explored, namely, a method that can be used to simulate 
real as well as imaginary quantum evolution under more relaxed 
conditions than those encountered in standard TEBD applications. 
This paper expands on the themes first addressed in 
reference \cite{ReslenRMF}, and can be seen as its continuation, 
especially in what concerns to the inclusion of
interaction and the treatment of non-unitary transformations.

Many of the efforts devoted during recent times to the
study of efficient simulation options have been motivated
by the remarkable advancement in cold atom experiments and
the possibility of probing fundamental theoretical models
in the laboratory \cite{Ronzheimer}. As a result, 
applications of numerical methods based on tensor states 
are getting increasingly common in descriptions of quantum 
gases in optical lattices \cite{Lacki,Piraud,Danshita,Reslen,Sorg}.
Tensor states display an assortment of properties that can
be exploited to absorb symmetries and implement transformations
selectively rather that on the whole Hilbert
space \cite{Feiguin,Singh,Clark}. Nevertheless, in the present work
tensor states are used as a tool that allow to put in practice
the design outlined throughout the analysis that follows.

Let us introduce the Bose-Hubbard Hamiltonian in one dimension 
as follows
\begin{equation}
\hat{H} = \sum_{j=1}^N \frac{U}{2} \hat{n}_{j} (\hat{n}_{j} - 1) - J \left ( \hat{a}_{j+1}^{\dagger} \hat{a}_{j} + \hat{a}_{j}^{\dagger} \hat{a}_{j+1} \right ) + \mu_j \hat{n}_{j},
\label{eq:1}
\end{equation}
being $\hat{n}_j = \hat{a}_{j}^{\dagger} \hat{a}_{j}$, while $\hat{a}_j$ and $\hat{a}_k^{\dagger}$ 
satisfy $[\hat{a}_j,\hat{a}_k^{\dagger}]=\delta_j^k$ and $[\hat{a}_j,\hat{a}_k]=0$ for $j,k=1,2,...,N$. 
Operators $\hat{a}_j$ and $\hat{a}_j^{\dagger}$ represent the $j$th mode. If
these modes are associated to Wannier functions with no significant
overlap in position \cite{Walters,Cazalilla}, as it is normally the case in applications
of the Bose-Hubbard model, then $j$ numbers a site
in a quantum chain. In order to allow for benchmarking against TEBD,
open boundary conditions are embraced, $\hat{a}_{N+1}^{\dagger}=\hat{a}_{N+1}=0$,
yet the validity of the central arguments here is not
tied to such a specific kind of boundary conditions. Moreover, 
periodic boundary conditions were used in \cite{ReslenRMF}.
The Bose-Hubbard model undergoes a continuous phase
transition from Mott insulator to Superfluid and
its phase diagram as well as its phenomenology has been
worked out in various contributions \cite{Fisher,Fazio,AMRey}.
\\
The standard version of TEBD makes use of the Trotter-Suzuki Expansion (TSE) \cite{Sanders} 
to decompose the evolution operator as a product of transformations involving only nearest 
neighbor sites. This is done by splitting the evolution operator
in two sub-evolutions, each generated by a part of the Hamiltonian
grouping every other term of the sum in (\ref{eq:1}). The approach
is efficacious, but it delimits the spectrum of
applications to Hamiltonians with nearest-neighbor hopping or
interaction, and open boundary conditions too. Here it is proposed to 
separate the Hamiltonian into single-particle
and many-particle operators, such as they are understood from a second-quantization perspective, 
and then write the evolution operator as a composition of
infinitesimal evolutions for each part. A second order split can
thus be considered as follows
\begin{equation}
e^{-i t_s \hat{H}} \approx e^{-i \frac{t_s}{2} \hat{H}_{MP}} e^{-i t_s \hat{H}_{SP}} e^{-i \frac{t_s}{2} \hat{H}_{MP}}, 
\label{eq:2}
\end{equation}
knowing that
\begin{gather}
\hat{H}_{SP} = \sum_{j=1}^N \left ( -\frac{U}{2} + \mu_j  \right ) \hat{n}_{j} - J \left ( \hat{a}_{j+1}^{\dagger} \hat{a}_{j} + \hat{a}_{j}^{\dagger} \hat{a}_{j+1} \right ), \label{eq:3} \\
\hat{H}_{MP} = \sum_{j=1}^N \frac{U}{2} \hat{n}_{j}^2. \label{eq:4}
\end{gather}
The only request being that the terms appearing in $\hat{H}_{MP}$,
i.e., the interaction, must be local, but $\hat{H}_{SP}$ can take any form as long as it 
remains single-body. Under the action of single-particle transformations, 
it is known that the mode operators inherit the properties normally 
assigned to states and in particular a form of linearity that in 
turn induces a form of unitariety which governs the coefficients of 
the mode operators. It is this unitariety that will be important
for the development of the current proposal and should be 
differentiated from the most common and general notion of unitariety.  

Independently of the formulation of the exponential split, the 
simulation protocols that have relevance in this study employ
the tensorial representation and the updating routines for one-
and two-site transformations on a quantum chain that
were first introduced in \cite{Vidal}. In coherence to this, the 
cost of a given computation is reported by the variable $\chi$, 
which establishes, in the shape of a polynomial, the maximum number 
of arithmetic operations necessary to calculate the action of an 
operator on at most two consecutive sites of the chain. 
It will be seen that all these elements
play a role in the conception and testing of the methodologies 
that are the focus of this review. 

In the next section the fundamentals of the technique, here referred
to in general as mode folding (MF), are explained. Then it is seen
how interaction is incorporated and how the whole proposal performs
against characteristic parameters and in comparison to TSE.
Subsequently, it is shown how to work with non-unitary transformations
and the particularities found in the calculation of ground states.
It is pointed out that different versions of the proposal can be
used together in cases where ground state and evolution are both
required, such as in studies of quench dynamics. Alternative approaches
and conclusions are presented in the last part.
\section{mode folding}
A solution of the Schrodinger equation for a system of identical particles
is given by
\begin{equation}
|\psi(t) \rangle = \prod_{q=1}^{N} \frac{\left( \hat{\alpha}_q^{\dagger} \right)^{n_q}}{\sqrt{n_q!}}|0\rangle,	
\label{eq:5}
\end{equation}
where $\hat{\alpha}_q^{\dagger}(t) = e^{-it\hat{H}} \hat{a}_q^{\dagger} e^{it\hat{H}}$ ($\hbar=1$).
$N$ is the number of modes and $M=\sum_{q=1}^N n_q$ is the total number of particles.
It follows from Eq. (\ref{eq:5}) that $|\psi(0) \rangle = \prod_{q=1}^{N} |n_q \rangle$. If only single-particle 
transformations are considered ($U=0$), then the
evolution modes behave linearly (for the sake of clarity, time dependencies are stressed in some of the subsequent 
expressions) 

\begin{equation}
\hat{\alpha}_q^{\dagger}(t) = c_{1,q}(t) \hat{a}_1^{\dagger} + c_{2,q}(t) \hat{a}_2^{\dagger} + ... + c_{N,q}(t) \hat{a}_N^{\dagger}. 
\label{eq:6}
\end{equation}
The coefficients in this expression depend essentially on $\hat{H}_{SP}$ through a closed
system of dynamical equations. Unitary operations on $|\psi(t)\rangle$ can be seen as
transformations that act simultaneously on every $\hat{\alpha}_q^{\dagger}$.
There are two types of such transformations that have applications in the
folding method. The first one affects mode operators individually and is used to
make the coefficients real. Such an effect is achieved by implementing the inverse of 
\begin{equation}
\hat{r}_k^{[l]}(t) = e^{ i \phi_{l,k}(t) \hat{a}_l^{\dagger} \hat{a}_l }, \text{ } \phi_{l,k}(t) = \arg c_{l,k} (t),
\label{eq:7}
\end{equation}
getting as a result $c_{l,k} \hat{a}_l^{\dagger} \rightarrow |c_{l,k}| \hat{a}_l^{\dagger}$.
When this is done consecutively for $l=1,2,...,N$ only real coefficients are left in the equivalent of
Eq. (\ref{eq:6}) with $q=k$. The $c_{l,k}$'s can then be redefined as the absolute values of the
original coefficients. The other type of transformation involves pairs of modes and is
given by the inverse of
\begin{gather}
\hat{R}_k^{[j+1,j]}(t) = e^{i \theta_{j,k}(t) \hat{J}_y^{[j+1,j]}}, \label{eq:8} \\ 
\hat{J}_y^{[j+1,j]} = \frac{1}{2i} \left( \hat{a}_{j+1}^{\dagger} \hat{a}_{j} - \hat{a}_{j}^{\dagger} \hat{a}_{j+1} \right). \label{eq:9}
\end{gather}
The transformation is essentially a rotation since its generator is a genuine quantum angular 
momentum. The procedure yields
\begin{gather}
 \left . \hat{R}_k^{[j+1,j]}\right .^{-1}   \left (  c_{j+1,q} \hat{a}_{j+1}^{\dagger} +  c_{j,q} \hat{a}_{j}^{\dagger} \right ) \hat{R}_k^{[j+1,j]}, \nonumber \\
=  c_{j+1,q}' \hat{a}_{j+1}^{\dagger} +  c_{j,q}' \hat{a}_{j}^{\dagger}, \label{eq:10} \\
c_{j+1,q}' = c_{j+1,q} \cos \left( \frac{\theta_{j,k}}{2} \right) - c_{j,q} \sin \left( \frac{\theta_{j,k}}{2} \right), \label{eq:11} \\
c_{j,q}' =   c_{j+1,q} \sin \left( \frac{\theta_{j,k}}{2} \right) + c_{j,q} \cos \left( \frac{\theta_{j,k}}{2} \right). \label{eq:12}
\end{gather}
The fact that the change on the coefficients is  unitary
is pivotal to the ensuing discussion. After making $c_{j+1,q}' = 0$ it follows
\begin{equation}
\tan \left( \frac{\theta_{j,q}}{2} \right) = \frac{c_{j+1,q}}{c_{j,q}}.
\label{eq:13}
\end{equation}
As a result, operator $\hat{a}_{j+1}^{\dagger}$ is neutralized in $\hat{\alpha}_q^{\dagger}$.
In order to see the complete action of these operations, 
let us propose a view in which the evolved modes appear stacked as shown by (\ref{eq:14}).
Transformations affect all the coefficients vertically aligned. In the
first part of the proposed protocol, all the coefficients in the first row
of the stack, i.e., those belonging to $\hat{\alpha}_1^{\dagger}$, are
stripped from their complex phases employing transformations of the first kind. Notice that once this has
been completed, all the coefficients in the stack have most likely changed, not just
the ones in the first row. The next step consists in making $\hat{a}_N^{\dagger}$
disappear from the first row by operating on the two columns at the left of the stack,
as indicated in (\ref{eq:14}), and then making $c_{N,1}'=0$. This cancellation 
technique is repeated systematically, advancing toward the right, until all the operators
except $\hat{a}_1^{\dagger}$ are eliminated from the first row, as indicated by 
(\ref{eq:15}) and (\ref{eq:16}). It is worth noticing that the last
of these operations entails the cancellation of all but the top $\hat{a}_1^{\dagger}$ in
the column at the right extreme of the stack, by virtue of the unitariety of all
previous transformations and the fact that the modes are orthonormal. At the end of this first series (or layer) of changes,
the first mode has been folded and the stack of operators appears just like (\ref{eq:17}). 
The second layer of operations is aimed at folding the second mode and has a very similar structure,
but it is carried without reaching the first mode, lest it unfolds. The
scheme goes on in an orderly manner until the stack is left with a different operator in
every level. At this point, it can be argued that the system has been returned to its
initial state independently of the distribution of $n_q$'s in Eq. (\ref{eq:5}).
\begin{gather}
\begin{array}{|c|c}  
\left . \hat{R}_1^{[N,N-1]} \right .^{-1}  &  \\ \cline{1-1}
c_{N,1}\hat{a}_N^{\dagger} + c_{N-1,1} \hat{a}_{N-1}^{\dagger} & + c_{N-2,1} \hat{a}_{N-2}^{\dagger}  \dots + c_{1,1} \hat{a}_1^{\dagger} \\ 
c_{N,2}\hat{a}_N^{\dagger} + c_{N-1,2} \hat{a}_{N-1}^{\dagger} & + c_{N-2,2} \hat{a}_{N-2}^{\dagger}  \dots + c_{1,2} \hat{a}_1^{\dagger} \\
\dots  &   \dots \\
c_{N,N}\hat{a}_N^{\dagger} + c_{N-1,N} \hat{a}_{N-1}^{\dagger} & + c_{N-2,N} \hat{a}_{N-2}^{\dagger} \dots + c_{1,N} \hat{a}_1^{\dagger} \\  
\end{array} \label{eq:14} \\
\begin{array}{c|c|c}  
& \left . \hat{R}_1^{[N-1,N-2]} \right .^{-1} &  \\ \cline{2-2}
 & c_{N-1,1}'\hat{a}_{N-1}^{\dagger} + c_{N-2,1} \hat{a}_{N-2}^{\dagger} &  \dots + c_{1,1} \hat{a}_1^{\dagger} \\ 
 c_{N,2}'\hat{a}_N^{\dagger} + & c_{N-1,2}'\hat{a}_{N-1}^{\dagger} + c_{N-2,2} \hat{a}_{N-2}^{\dagger} &  \dots + c_{1,2} \hat{a}_1^{\dagger} \\
\dots & \dots & \dots \\
 c_{N,N}'\hat{a}_N^{\dagger} + & c_{N-1,N}'\hat{a}_{N-1}^{\dagger} + c_{N-2,N} \hat{a}_{N-2}^{\dagger} &  \dots + c_{1,N} \hat{a}_1^{\dagger} \\  
\end{array} \label{eq:15} \\ 
\vdots \nonumber \\
\begin{array}{c|c|}  
& \left . \hat{R}_1^{[2,1]} \right .^{-1}  \\ \cline{2-2}
   & c_{2,1}'\hat{a}_{2}^{\dagger} + c_{1,1} \hat{a}_{1}^{\dagger}  \\ 
 c_{N,2}'\hat{a}_N^{\dagger} + \dots + c_{3,2}''\hat{a}_{3}^{\dagger} + & c_{2,2}'\hat{a}_{2}^{\dagger} + c_{1,2} \hat{a}_{1}^{\dagger}  \\
\dots & \dots  \\
 c_{N,N}'\hat{a}_N^{\dagger} + \dots + c_{3,N}''\hat{a}_{3}^{\dagger} + & c_{2,N}'\hat{a}_{2}^{\dagger} + c_{1,N} \hat{a}_{1}^{\dagger}  \\  
\end{array} \label{eq:16} \\
\nonumber \\
\begin{array}{cc}  
   & \hat{a}_{1}^{\dagger}  \\ 
 c_{N,2}'\hat{a}_N^{\dagger} + \dots + c_{3,2}''\hat{a}_{3}^{\dagger} + c_{2,2}''\hat{a}_{2}^{\dagger}  &   \\
\dots &  \\
 c_{N,N}'\hat{a}_N^{\dagger} + \dots + c_{3,N}''\hat{a}_{3}^{\dagger} + c_{2,N}''\hat{a}_{2}^{\dagger}  &   \\  
\end{array} \label{eq:17}
\end{gather}
Having determined the set of transformations required by this folding mechanism, 
it is possible to reassemble the evolved state implementing the sequence in reverse
order. All of this can be mathematically synthesized as 
\begin{equation}
| \psi(t) \rangle = \prod_{k=N}^1 \left ( \prod_{l=k}^{N} \hat{r}_k^{[l]}(t) \prod_{j=k<N}^{N-1} \hat{R}_k^{[j+1,j]}(t) \right ) | \psi(0) \rangle.
\label{eq:18}
\end{equation}
However, Eq. (\ref{eq:18}) is not the only possibility. In particular, if the
stack is folded starting by $\hat{\alpha}^{\dagger}_N$, it is found instead 
\begin{equation}
| \psi(t) \rangle = \prod_{k=1}^N \left ( \prod_{l=N}^{N-k+1} \hat{r}_k^{[l]} \prod_{j=N-1}^{N-k+1<N} \hat{R}_k^{[j+1,j]} \right ) | \psi(0) \rangle.
\label{eq:19}
\end{equation}
Henceforth this latter approach will be referred to as the inverse MF, 
in order to differentiate it from the former scheme, or ``normal'' MF. 
Although both forms are in principle equivalent, it will be seen that
their numerics might display different accuracy.
\section{The inclusion of interaction and the case of rapidly decaying coefficients}
\begin{figure*}
\centering
\includegraphics[width=0.233\textwidth,angle=-90]{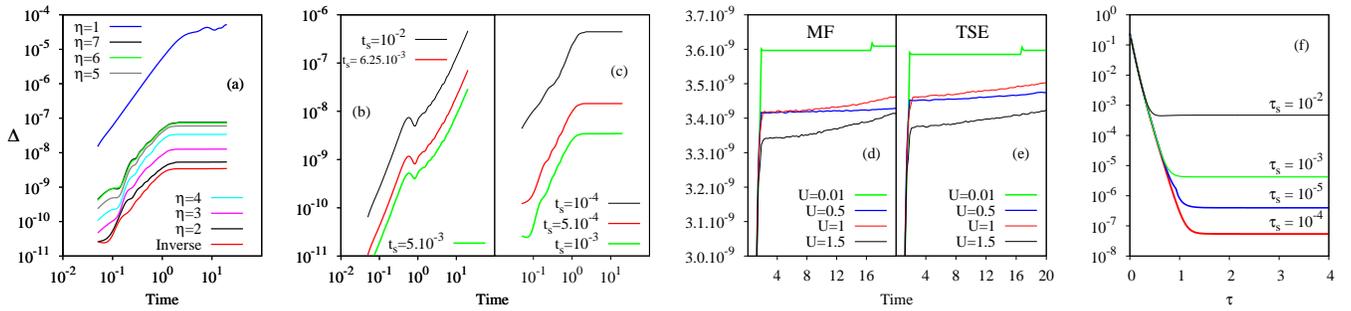}
\caption{Numerical error delivered by MF under 
different circumstances in boson chains of size $N=M=8$. (a) These 
calculations correspond to Hamiltonian (\ref{eq:1}) with $U=1$, 
$J=1$, $\mu=0$. In addition, $t_s=10^{-3}$ and $\chi =105$.
(b)-(c) Inverse MF for different time slices. The simulated systems 
correspond to $U=1$, $J=1$, $\mu=0$,  and $\chi =105$. 
(d)-(e) Error comparison between second order implementations of 
inverse MF and TSE for different $U$. In both cases $J=1$, 
$\mu=0$, $t_s=10^{-3}$ and $\chi =105$. 
(f) $\Delta = 1 - |\langle \psi_G | \psi(\tau) \rangle|^2$ 
for various values of $\tau_s$ in ground state calculations  
for $U=10$, $J=1$, $\mu=0$ and $\chi=\orderof (10)$. 
In all six graphs the initial state displays one boson at each site.} 
\label{fig1}
\end{figure*}
Since $| \psi(0) \rangle$ in Eqs. (\ref{eq:18}) and (\ref{eq:19}) can be any Fock state, 
it is possible to extend the formalism by writing an arbitrary initial state
as a superposition of Fock states and then using the fact that the quantum
evolution operator acts linearly on such a superposition. In order to
incorporate interaction effects, the evolution operator is split, as
in Eq. (\ref{eq:2}), into single-particle and many-particle
step evolutions. The tensorial representation of the state is obtained applying
the unitary transformations related to MF, which account for the single-particle
step, as well as the many-particle step 
using the updating protocols demonstrated in Ref. \cite{Vidal}. Furthermore, 
because the interaction terms are assumed to be local, the latter step does
not alter $\chi$, i.e., it does not produce changes in the size
of the tensorial representation. To test the proposal, the obtained matrix product state (MPS) 
is passed to the Fock basis and then compared to benchmark results
found from the diagonalization of the corresponding Hamiltonian in the Fock basis.
In order to manage the error incurred by splitting the
evolution operator, $t_s$ must be made small compared to the eigenvalues
of the Hamiltonian. As a consequence, the distribution of coefficients
 in (\ref{eq:14}) is highly sparse, with
most of the coefficients falling below machine accuracy. The
exception being diagonal elements, which remain close to one, and their neighbors. 
As small coefficients can be a source of numerical errors,
it is convenient to simply drop them and fold what is left of the stack.
This leads to a slightly different version of Eq. (\ref{eq:18}), namely
\begin{equation}
| \psi(t + t_s) \rangle = \prod_{k=N}^1 \left ( \prod_{l=k}^{k+\eta<N} \hat{r}_k^{[l]} \prod_{j=k<N}^{k+\eta-1<N-1} \hat{R}_k^{[j+1,j]} \right ) | \psi(t) \rangle,
\label{eq:20}
\end{equation}
where $\eta$ is the largest number of non-vanishing coefficients on either side 
of the stack diagonal. In order to measure the difference between the
results delivered by MF and other reference results, the following
error measure is introduced $\Delta = 1 - |\langle \psi' | \psi \rangle|^2$,
where $| \psi \rangle'$ comes from a tensorial representation and 
$| \psi \rangle$ from a standard matricial analysis.
It can be seen in figure \ref{fig1}(a) that different versions of MF
produce results with distinct accuracies. It can also be seen that
the best estimation is given by the inverse MF, although the 
difference with the next best estimation is rather marginal.
The observed optimal parameter is $\eta=2$, but in general
the appropriate $\eta$ depends on the distribution of coefficients
in (\ref{eq:14}). For small values of $t_s$ the distribution is tight
and the optimal $\eta$ is short. Likewise, $\eta$ should grow with increasing $t_s$.
It is empirically seen that the best performance is achieved when the smallest 
considered coefficient is approximately of the same order of magnitude 
than the square root of machine accuracy. The reduction in correctness 
observed when more coefficients are included occurs because substantial error
is transmitted to some folding angles calculated using coefficients with
an insufficient number of significant figures at the beginning of each layer of
calculations. Contrariwise, no such a tendency is observed in the inverse 
scheme since, even though poorly estimated coefficients are retained, the 
folding routine starts by the biggest coefficients on each layer. This makes
inverse MF reliable, although it requires the inclusion of all the folding
operations. Normal folding taking less coefficients is numerically more efficient,
but a little bit less exact.
Figures \ref{fig1}(b) and \ref{fig1}(c) show the characteristic behavior of 
the error as a function of time for different values of the time slice. 
For relatively large $t_s$, $\Delta$'s growth is essentially polynomial.
As $t_s$ diminishes, the corresponding error curves scale down until for an optimal
time slice the error starts showing saturation. As $t_s$ diminishes further, the
error scales up but the saturation profile remains. It can therefore be
argued that while the theoretical error is gradually suppressed with decreasing $t_s$, the
computing error caused by division by small numbers augments.
Similarly, the optimal value $\chi=105$ seems to be independent
of the values given to $\eta$ and $t_s$. However, the dependence
of the error with respect to $\chi$ is rather characteristic.
Making $\chi$ slightly smaller than its optimal value
produces exponentially growing deviations from the actual state at very
early times. This contrasts
with the behavior of $\Delta$ in simulations where although the time
slice has been chosen well above its optimal value, $\chi$ has
remained at or above its equilibrium value. In such cases the
error growth is polynomial and the simulation can be sustained
for quite longer intervals. In this sense it can be said that a right estimation
of $\chi$ is more important for numerics than the chosen $t_s$. This feature is encountered 
in both MF and TSE simulations.
The comparative graphs depicted in figures \ref{fig1}(d) and \ref{fig1}(e) suggest that 
the difference in terms of deviations from the correct state between MF and TSE is
minimal over a wide range of values of the ratio $U/J$. Since
it has been shown in Ref. \cite{ReslenRMF} that folding alone delivers
results with a tolerance of the order of machine accuracy, it 
follows that the observed error is mostly due to the splitting of the
evolution operator into single- and many particle parts and that 
such an error is comparable to the one produced by TSE. 
In terms of efficiency, because both approaches seem to require the
same $\chi$, the determinant factor is the
number of transformations effectuated on pairs of modes per loop in a
computer routine, i.e., the number of updates of the tensorial representation
necessary to advance the state a time $t_s$. In a second order version
such a number is approximately $3N/2$ for TSE and $N \eta - \eta (\eta+1)/2$ for MF.
The difference can be understood in terms of the cost associated to
the inclusion of more hopping terms in MF, in contrast to the less-strict
neighbor approach dictated by TSE.
\section{Non-unitary transformations and the calculation of ground states}
\begin{figure}
\centering
\includegraphics[width=0.32\textwidth,angle=-90]{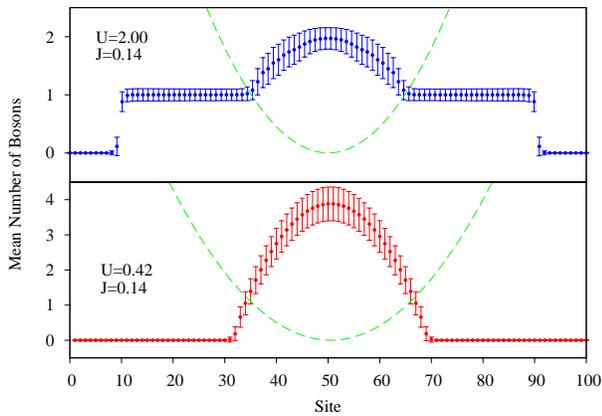}
\caption{Mean number of particles and fluctuations (error bars) for the ground state
of a boson chain with $N=M=100$ and open boundary conditions. In both cases the 
confining potential is $\mu_j = 0.0046(j-50.5)^2$ (dashed line). Additionally, 
$\tau_s = 10^{-5}$ and $\chi = \orderof(10)$.} 
\label{fig5}
\end{figure}

\begin{figure}
\centering
\includegraphics[width=0.6\textwidth,angle=-90]{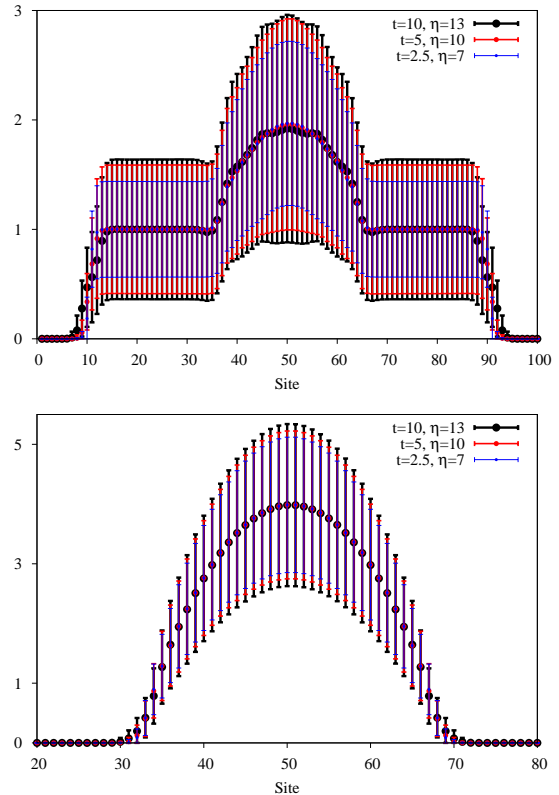}
\caption{Mean number of particles and fluctuations after a quench. 
Once interaction and trapping have been shut down, $U=\mu=0$, 
the dynamics is generated only by the kinetic term $J=0.14$. The
simulation constant is $\chi = \orderof(10^3)$. The corresponding 
states at $t=0$ are shown in figure \ref{fig5}. Mean values do not 
change much, rather, it is fluctuations that capture the state 
evolution.} 
\label{fig6}
\end{figure}

Ground states play a prominent role in the study of quantum systems and 
it is known that they usually display less entanglement than dynamical
states. As a consequence, the associated simulation cost in terms of
tensor states is manageable to a great extent. Simulation protocols
that rely on MPS can be formulated as variational methods. This is 
the case of DMRG, where the ground state is found as the network that
minimizes the energy. Another approach is to realize the limit
\begin{equation}
|\psi_G \rangle = \lim_{\tau \rightarrow \infty} \frac{e^{-\tau \hat{H}} |\psi_0\rangle}{||e^{-\tau \hat{H}} |\psi_0\rangle||},
\label{eq:21}
\end{equation}
as an iteration of an infinitesimal imaginary-time evolution of length $\tau_s$. Such
is the strategy followed in the context of TEBD and it is also the
approach adopted here. The crux of the problem is that imaginary-time
evolutions are not unitary, ergo the foundation of the
MPS updating method \cite {Orus} and the folding technique is compromised. To deal
with the issue it is important, on the one hand, to use a sufficiently small 
``imaginary'' time slice $\tau_s$, so that the corresponding advancement is 
close enough to unitary and the updating protocols work. On the other hand, it is 
necessary to find a folding protocol that is not as heavily dependent on the unitariety
of the evolution operation. Moreover, the protocol itself must integrate genuine 
non-unitary transformations so as to account for the non-unitariety of 
the whole transformation. In accordance with these premises, let us 
introduce  
\begin{equation}
\hat{Q}^{[j+1,j]} = e^{\epsilon_j \hat{J}_{xz}^{[j+1,j]}},\label{eq:25} 
\end{equation}
where,
\begin{gather}
\hat{J}_{xz}^{[j+1,j]} = \cos \varphi_j \hat{J}_x^{[j+1,j]} + \sin \varphi_j \hat{J}_z^{[j+1,j]},\label{eq:26} \\
\hat{J}_x^{[j+1,j]} = \frac{1}{2} \left( \hat{a}_{j+1}^{\dagger} \hat{a}_{j} + \hat{a}_{j}^{\dagger} \hat{a}_{j+1} \right),\label{eq:27} \\
\hat{J}_z^{[j+1,j]} = \frac{1}{2} \left( \hat{a}_{j+1}^{\dagger} \hat{a}_{j+1} + \hat{a}_{j}^{\dagger} \hat{a}_{j} \right).\label{eq:28}
\end{gather}
Transformation (\ref{eq:25}) depends on $\epsilon_j$, which is supposed to
be small, and also on $\varphi_j$, which can take any value in $[0,2\pi]$.
The operation is defined in terms of the quantum angular momenta $\hat{J}_x$  
and $\hat{J}_z$ but it is not strictly a rotation. The action is confined to
the $j$th and $j+1$th modes and the effect of the inverse on the coefficients of
the $q$th row is
\begin{gather}
c_{j+1,q}' = - c_{j,q} \cos \varphi_j  \sinh \left( \frac{\epsilon_{j}}{2} \right) + \nonumber \\   
c_{j+1,q} \left (  \cosh \left( \frac{\epsilon_{j}}{2} \right) - \sin \varphi_j \sinh \left( \frac{\epsilon_{j}}{2} \right )  \right ), \label{eq:29} \\
c_{j,q}' =  
-c_{j+1,q} \cos \varphi_j  \sinh \left( \frac{\epsilon_{j}}{2} \right) 
+ \nonumber \\
c_{j,q} \left (  \cosh \left( \frac{\epsilon_{j}}{2} \right) + \sin \varphi_j \sinh \left( \frac{\epsilon_{j}}{2} \right )  \right ). \label{eq:30}
\end{gather}
Demanding that both $c_{j+1,q}' = 0$ and $c_{j,q+1}' = 0$, it results
\begin{gather}
\tan \varphi_j = \frac{c_{j+1,q} c_{j+1,q+1} - c_{j,q} c_{j,q+1}}{2 c_{j+1,q} c_{j,q+1}}, \label{eq:31}\\
\tanh \left( \frac{\epsilon_{j}}{2} \right) = \frac{ c_{j+1,q}  }{ c_{j+1,q} \sin \varphi_j + c_{j,q} \cos \varphi_j}.\label{eq:32}
\end{gather}
It can be seen that the hyperbolic tangent is well defined
as long as $|c_{j+1,q}| \ll |c_{j,q}|$ and $|c_{j,q+1}| \ll |c_{j+1,q+1}|$,
except if $c_{j+1,q} = 0$ or $c_{j,q+1} = 0$. These consistency conditions are 
met if $\tau_s$ is sufficiently short but non-vanishing. In particular, $\tau_s$ 
can be set to a value for which $\eta=1$, yielding the operator stack shown
in (\ref{eq:22}). In this context the coefficients are most likely real and positive, if 
they are not, they can be adjusted using unitary operations as previously discussed.
A two-mode transformation is applied on the right-extreme of the stack,
resulting in the simultaneous cancellation of $\hat{a}_2^{\dagger}$ in the first row and
$\hat{a}_1^{\dagger}$ in the second row \cite{comment3}, leaving the stack as indicated
in (\ref{eq:23}). The lack of unitariety of (\ref{eq:23}) means coefficient
$c_{1,1}'$ can be hardly equal to one, hence it is normalized using the inverse of 	 
$\hat{q}^{[j]} = e^{\delta_j \hat{a}_j^{\dagger} \hat{a}_j}$, with $\delta_j = \log c_{j,j}''$,
when $j=1$. Thereafter the first mode is folded and the cancellation-and-normalization sequence 
starts over again as depicted by (\ref{eq:23}) and (\ref{eq:24}).
\begin{gather}
\begin{array}{r|l|}  
& \multicolumn{1}{c|} {\left .  \hat{Q}^{[2,1]} \right .^{-1} } \\ \cline{2-2}
&  c_{2,1} \hat{a}_{2}^{\dagger} + c_{1,1} \hat{a}_1^{\dagger} \\ 
c_{3,2} \hat{a}_{3}^{\dagger} + & c_{2,2} \hat{a}_{2}^{\dagger} + c_{1,2} \hat{a}_1^{\dagger} \\
c_{4,3} \hat{a}_{4}^{\dagger} + c_{3,3} \hat{a}_{3}^{\dagger} +  & c_{2,3} \hat{a}_{2}^{\dagger} \\
c_{5,4} \hat{a}_{5}^{\dagger} + c_{4,4} \hat{a}_{4}^{\dagger} + c_{3,4} \hat{a}_{3}^{\dagger} \text{\hspace{0.3cm}}  &  \\
\multicolumn{1}{c|}{\iddots}  &    \\
\end{array} \label{eq:22} \\
\begin{array}{r|c|c|}  
& \left . \hat{Q}^{[3,2]} \right .^{-1} & \left . \hat{q}^{[1]} \right .^{-1}  \\ \cline{2-3}
&  & c_{1,1}' \hat{a}_1^{\dagger} \\ 
& c_{3,2} \hat{a}_{3}^{\dagger} + c_{2,2}' \hat{a}_{2}^{\dagger}  & \\
c_{4,3} \hat{a}_{4}^{\dagger} +   & c_{3,3} \hat{a}_{3}^{\dagger} + c_{2,3}' \hat{a}_{2}^{\dagger} & \\
c_{5,4} \hat{a}_{5}^{\dagger} + c_{4,4} \hat{a}_{4}^{\dagger} +   &\multicolumn{1}{l|}{c_{3,4} \hat{a}_{3}^{\dagger}} &  \\
\multicolumn{1}{c|}{\iddots}  & &    \\
\end{array} \label{eq:23} \\ 
\begin{array}{r|c|c|c}  
& \left . \hat{Q}^{[4,3]} \right .^{-1}  & \left . \hat{q}^{[2]} \right .^{-1} &  \\ \cline{2-3}
&  &  & \text{\hspace{0.48cm}} \hat{a}_1^{\dagger} \\ 
&  & c_{2,2}'' \hat{a}_{2}^{\dagger}  & \\
& c_{4,3} \hat{a}_{4}^{\dagger} +  c_{3,3}' \hat{a}_{3}^{\dagger}  & & \\
c_{5,4} \hat{a}_{5}^{\dagger} +  & c_{4,4} \hat{a}_{4}^{\dagger} + c_{3,4}' \hat{a}_{3}^{\dagger}  & &  \\
\multicolumn{1}{c|}{\iddots}  & & &  
\end{array} \label{eq:24}
\end{gather}
Gathering all the involved steps and putting them in reverse order, 
the advanced state can be written as
\begin{equation}
| \psi(\tau + \tau_s) \rangle = \left ( \prod_{j=N-1}^1 \hat{Q}^{[j+1,j]} \hat{q}^{[j]} \right ) \hat{q}^{[N]} | \psi(\tau) \rangle.
\label{eq:34}
\end{equation}
Furthermore, interaction mechanisms can be incorporated using the split-operator method,
analogously to the case of normal evolution. Beginning with a state that has an overlap with 
the ground state, the whole set of operations is applied systematically until a
convergence criterion is satisfied. Figure \ref{fig1}(f) depicts the convergence
behavior as captured by measuring the fidelity to the actual ground state as
a function of $\tau$. For the cases studied, convergence is steady but 
tolerance depends on $\tau_s$. As seen for real time calculations, there exists an 
optimal slice that minimizes the simulation error.
The test simulations displayed in figure \ref{fig5} show a system of 
interacting bosons subject to a confining potential under the regimes of Mott 
Insulator (top) and Superfluid (bottom). Both phases can be realized in experiments
of cold atoms in optical lattices with the parameters used in the plots of figure \ref{fig5}. 
These outcomes coincide qualitatively with the results 
obtained in \cite{Reslen_thesis} using TSE in chains with the same set of 
parameters and size. Of relevance is the fact that, as long as $\chi$ is optimal, the symmetry of the 
state is correctly reproduced by the folding method in spite of the relatively large
number of operations correlating modes that are distant. All this reaffirms
that little error is genrated by the folding process alone, even in the
presence of interaction.
One advantage of MF is that real- and imaginary-time versions can
be used in conjunction to simulate quench dynamics where the initial
state is set to the ground state of a Hamiltonian with dominant 
interaction. This is especially useful when the evolution is governed 
by a single-particle Hamiltonian. Figure \ref{fig6} exemplify this 
approach using the ground states featured in figure \ref{fig5} as 
initial states. After interaction and confining potential have been 
switched off, atoms evolve freely toward a fluctuation-dominated 
phase. In the time span considered, the atom cloud does not expand 
significantly beyond its initial domain, instead, there is an 
increase in hopping across the whole extent of the cloud. As longer
times are addressed, the simulation cost soars. 

The techniques introduced in this work make use of the basic idea of folding 
of modes to develop numerical routines that calculate the quantum state. Although
each application is different, the essence of every method is the same. Certainly,
the folding idea is potentially versatile and may give rise to additional 
simulation protocols. To give an example of another application, let us conceive
the dynamical modes as a superposition of the Hamiltonian eigenvectors as
follows
\begin{equation}
\left (
\begin{array}{c}
\alpha_1^{\dagger}(t_s) \\
\alpha_2^{\dagger}(t_s) \\
\alpha_3^{\dagger}(t_s) \\
\vdots
\end{array}
\right )
=
\hat{b}_1^{\dagger} e^{-i t_s E_1}
\left (
\begin{array}{c}
E
_{11} \\
E_{21} \\
E_{31} \\
\vdots
\end{array}
\right ) +
\hat{b}_2^{\dagger} e^{-i t_s E_2}
\left (
\begin{array}{c}
E_{12} \\
E_{22} \\
E_{32} \\
\vdots
\end{array}
\right ) + \dots
\label{eq:35}
\end{equation}
so that
\begin{equation}
\hat{b}_q^{\dagger} = \hat{a}_1^{\dagger} E_{1q} + \hat{a}_2^{\dagger} E_{2q} + \hat{a}_3^{\dagger} E_{3q} + \dots
\label{eq:36}
\end{equation}
Initially, making $t_s=0$, the superposition lacks the exponential terms.
Therefrom, the first move is to fold $\hat{b}_1^\dagger$ using unitary transformations 
until only $\hat{a}_1^\dagger$ remains. Without further action, $\hat{a}_1^\dagger$
is automatically expelled from the updated versions of all the other $\hat{b}$'s. Next, 
the single-particle operation $e^{-i t_s E_1 \hat{a}_1^{\dagger} \hat{a}_1}$ is applied. 
The inverses of the transformations employed to fold $\hat{b}_1^\dagger$ are then 
executed in reverse order. This brings $\hat{b}_1^\dagger$ back into the superposition, 
but accompanied by the term $e^{-i t_s E_1}$, leaving the other $\hat{b}$'s
as they were at the beginning. An analogous sequence is then applied, focusing
this time on $\hat{b}_2^\dagger$, in order to incorporate $e^{-i t_s E_2}$. The 
rest of the protocol ensues in a logical manner until all the exponentials are
introduced. The transformations are then collected and put together to 
assemble an evolution operator that can be implemented in terms of tensor 
states. A very similar scheme can be formulated to handle non-unitary transformations.
The problem is that the corresponding numerics is not very stable owing to 
the recurrent folding-unfolding sequence, but the proposal is perfectly valid.
Additional folding schemes will be reported in future contributions.

\section{Conclusions}

Numerical techniques with applications to interacting systems 
have been introduced and probed. The proposals are developed in the
context of local interactions and are based on the idea of mode folding and
on the efficient use of tensor states. Real as well as imaginary
time implementations of the evolution operator are discussed,
elaborating on the handling of each particular case. 
The method compares well with TSE in terms of accuracy and
simulation time. However, its potential mainly resides, on the one hand, 
in the possibility of managing hopping of arbitrary scope, and on the other 
hand, in the perspective that it offers 
about the role of single-body- and many-body structures in the 
calculation of a quantum state. As an extension, it would be interesting to
consider the prospects of folding of fermionic modes. As it is known,
the sign-problem prevents the effective use of TSE.
It remains to be seen whether the alternative path taken by MF
somehow allows to circumvent the adverse effects of fermion algebra.


\begin{thebibliography}{}
\bibitem{comment3} In addition, this operation turns on $\hat{a}_1^{\dagger}$ in the 
third row, but with a coefficient that is below machine accuracy.
\end{thebibliography}
\end{document}